\documentclass[structabstract]{aa}  
\usepackage{graphicx}
\usepackage[applemac]{inputenc}
\usepackage{txfonts}
\usepackage{natbib}
\bibpunct{(}{)}{;}{a}{}{,} 
\begin{document}
   \title{Solar-like oscillations in the G9.5 subgiant $\beta$ Aquilae\thanks{Based on observations collected at the Telescopio Nazionale Galileo (TNG), La Palma, Spain.}}
   \author{E. Corsaro\inst{1}\fnmsep\inst{2}\fnmsep\thanks{\email{eco@oact.inaf.it}},
          F. Grundahl\inst{3},
          S. Leccia\inst{4}, 
          A. Bonanno\inst{2},
          H. Kjeldsen\inst{3}
          \and
          L. Paternò\inst{1}\fnmsep\inst{2}
          }

   \institute{\inst{1}Department of Physics and Astronomy, Astrophysics Section, University of Catania,
              Via S. Sofia 78, I-95123 Catania, Italy\\
             \inst{2}I.N.A.F. - Astrophysical Observatory of Catania, Via S. Sofia 78, I-95123 Catania, Italy\\
             \inst{3}Department of Physics and Astronomy, Aarhus University, DK-8000 Aarhus C, Denmark\\
             \inst{4}I.N.A.F. - Astronomical Observatory of Capodimonte, Salita Moiariello 16, I-80131 Napoli, Italy
             }

   \date{Received ; Accepted 4 Oct 2011}
  \abstract
   {An interesting asteroseismic target is the G9.5 IV solar-like star $\beta$ Aql. This is an 
   ideal target for asteroseismic investigations, because precise astrometric 
   measurements are available from \textit{Hipparcos} that greatly help in constraining the theoretical interpretation of the results. The star was 
   observed during six nights in August 2009 by means of the high-resolution échelle spectrograph
   SARG operating with the TNG 3.58 m Italian telescope (Telescopio Nazionale Galileo) 
   on the Canary Islands, exploiting the iodine cell technique.}
   {We present the result and the detailed analysis of high-precision radial velocity measurements, where the possibility of detecting individual p-mode frequencies for the first time and deriving their corresponding asymptotic values will be discussed.}
   {The Fourier analysis technique based on radial velocity time series and the fitting of asymptotic relation to the power spectrum were the main tools used for the detection of the asteroseismic parameters of the star.}
   {The time-series analysis carried out from $\sim$ 800 collected spectra shows the typical p-mode frequency pattern with a maximum centered at $416$ $\mu$Hz. In the frequency range 300 - 600 $\mu$Hz we identified for the first time six high signal-to-noise ratio (S/N $\gtrsim 3.5$) modes with $\ell = 0,2$ and $11 < n < 16$ and three possible candidates for mixed modes ($\ell = 1$), although the p-mode identification for this type of star appears to be quite difficult owing to a substantial presence of avoided crossings. The large frequency separation and the surface term from the set of identified modes by means of the asymptotic relation were derived for the first time. Their values are $\Delta \nu = 29.56 \pm 0.10$ $\mu$Hz and $\epsilon = 1.29 \pm 0.04$,  consistent with expectations. The most likely value for the small separation is $\delta\nu_{02} = 2.55 \pm 0.71$ $\mu$Hz. }
   {}
   
    \authorrunning{Corsaro et al.}

   \maketitle
%

\section{Introduction}
The search for solar-like oscillations \citep[see][for a summary]{CD04,Elsworth04} in main-sequence and subgiant stars showed a tremendous growth in the last decade \citep[for reviews see e.g.][]{Bedding03,Bedding06}, especially by means of the photometric space-based missions \textit{CoRoT} \citep{CoRoT06,CoRoT08} and \textit{Kepler} \citep{Borucki10,Koch10}. The latter in particular is presently providing an enormous amount of high-quality asteroseismic data \citep[e.g. see][for more details on the KAI - \textit{Kepler} Asteroseismic Investigation]{Gilliland10a}. Photometric studies of a large number of solar-type stars are fundamental for statistical investigations of intrinsic stellar properties and for testing theories of stellar evolution \citep[e.g.][]{Chaplin11Sci}. 
However, as is known from the theory of solar-like oscillations, high-precision Doppler shift measurements are more effective for detecting p modes of higher angular degrees. At the present time the échelle spectrometers such as CORALIE, HARPS, UCLES, UVES, SOPHIE and SARG, attaining high-precision radial velocity (RV) measurements \citep{Geoffrey92}, offer a way to detect solar-like oscillations in bright asteroseismic targets \citep[e.g. see][for a review on ground-based campaigns]{Bedding08}. 
The spectroscopic approach to the detection of solar-like oscillations will also be used in the  ground-based SONG project \citep{SONG} in the near future.
 
In this paper we report the detection of excess of power and perform a detailed analysis based on the radial velocity measurements for the evolved subgiant star $\beta$ Aql (HR 7602, HD 188512, HIP 98036). This star has $V =  3^m.699 \pm 0^m.016$ \citep[UBV photometric measurements from][]{Oja86}, spectral type G9.5 IV \citep{Gray06}, distance $13.70 \pm 0.04$ pc derived from the \textit{Hipparcos} parallax $\pi = 73.00 \pm 0.20$ mas \citep{HIPPARCOS}, $T_\mathrm{eff} = 5160 \pm 100$ K \citep{Luck05}, $\log g = 3.79 \pm 0.06$ \citep{Valenti05} and $[Fe/H] = -0.17 \pm 0.07$ \citep{Fuhrmann04}. 
The radius $R = 3.05 \pm 0.13$ $R_{\odot}$ was measured by means of Long Baseline Interferometry \citep{Nordgren1999} that is listed in the CHARM2 catalogue for high angular resolution measurements \citep{Richichi05}. 
An excess of power in the power spectrum (PS) of $\beta$ Aql data acquired with HARPS was already found by \cite{Kjeldsen08baql}, who estimated a mean large separation $\Delta \nu = 30$ $\mu$Hz from stellar parameters.
In Section~\ref{sc:sec2} we describe the observations, the data reduction and the RV extraction developed for this star. The Fourier analysis and the mode characterization that led us to derive the asteroseismic parameters are presented in Section~\ref{sc:sec3}. In Section~\ref{sec:theor} we discuss the scaled mass, mode amplitude, and frequency of maximum power of the star, where a comparison with expectations and a global list of stellar parameters also derived by means of the SEEK package \citep{Quirion10}.

\section{Observation and data reduction}
\label{sc:sec2}
The data were acquired with observations carried out during six nights (2009 August 5-11) by means of the high-resolution cross-dispersed échelle spectrograph SARG \citep{Gratton01,Claudi09} mounted on the 3.58 m Italian telescope TNG at the La Palma observing site. 
    SARG operates in both single-object and long-slit (up to 26”) observing modes and covers a spectral wavelength range from 370 nm up to about 1000 nm, with a resolution ranging from $R = 29000$ up to $R = 164000$. Our spectra were obtained at $R = 164000$ in the wavelength range 462-792 nm. The calibration iodine cell works only in the blue part of the spectrum (500-620 nm), which was used for measuring Doppler shifts. During the observing period we collected spectra with a signal-to-noise ratio (S/N) varying from 150 to 300, a typical exposure of $\sim$ 150 s and a sampling time of $\sim$~190 s. A total of 828 spectra were collected with the following distribution over the six nights: 91, 133, 110, 146, 163, 185. The spectra were then reduced and calibrated in wavelength with a Th-Ar lamp, using standard tasks of the IRAF package facilities. No flat-fielding was applied to the spectra because of a degradation of the S/N level.

\subsection{Radial velocity measurements} 
\label{sc:sec2-1}
The RV measurements were determined by means of the iSONG code, developed for the SONG project \citep{SONG}. iSONG models the instrumental profile (PSF), stellar, and iodine cell spectra to measure Doppler shifts. The observed spectrum was fitted with a reconstructed one by using a 
 convolution between the oversampled stellar template, the very high-resolution iodine cell spectrum, and the measured spectrograph instrumental profile. Essential to this process are the template spectra of $\beta$ Aql taken with the iodine cell removed from the beam, and the iodine cell itself superimposed on a rapidly rotating B-type star, the same for all the measurements. The B-type star spectra allow us to construct the instrumental profile, 
while the $\beta$ Aql spectra acquired without the iodine cell create the original stellar template when deconvolved with the PSF \cite[see][for a detailed explanation of the method]{Butler96}. The velocities were  corrected to the solar system barycenter (Hrudkovà et al. 2006) and no other corrections, such as decorrelation or filtering by removing polynomial fits to the time-series, were applied. 

iSONG provides an estimate of the uncertainty in the velocity measurements, $\sigma_i$; these values 
were derived from the scatter of the velocities measured from many ($\simeq 650$), small ($\simeq 1.8$  $\AA$) segments (chunks) of the échelle spectrum. To include
the accuracy of the measurements in a weighted Fourier analysis of the data, we firstly verified that these $\sigma_i$ reflected the noise properties of the RV measurements and their Fourier transform, following a slightly modified \cite{Butler04} approach.
First we considered that the variance deduced from the noise level $\sigma_{\mathrm{amp}}$ in the amplitude spectrum has to satisfy the relationship
\begin{equation}
      \sigma^{2}_{\mathrm{amp}} \sum^{N}_{i=1} \sigma^{-2}_i = \pi
   \end{equation}
according to \cite{Kjeldsen92} and \cite{Kjeldsen95}. This procedure yields new RV uncertainties scaled down by  a factor of 1.51. Then we performed the following three steps. (i) The high-frequency noise in the power spectrum (PS), well beyond the stellar signal ($> 1000$ $\mu$Hz), reflects the properties of the noise in the RV data, and because we expected that the oscillation signal is the dominant cause of variations in the velocity time series, we need to remove it to analyze the noise. This we made iteratively by finding the strongest peak in the PS of the velocity time-series and subtracting the corresponding sinusoid from the time-series data \citep[see][Section 4.3]{Frandsen95}. 
\begin{figure}
   \centering
   \includegraphics[width=9cm]{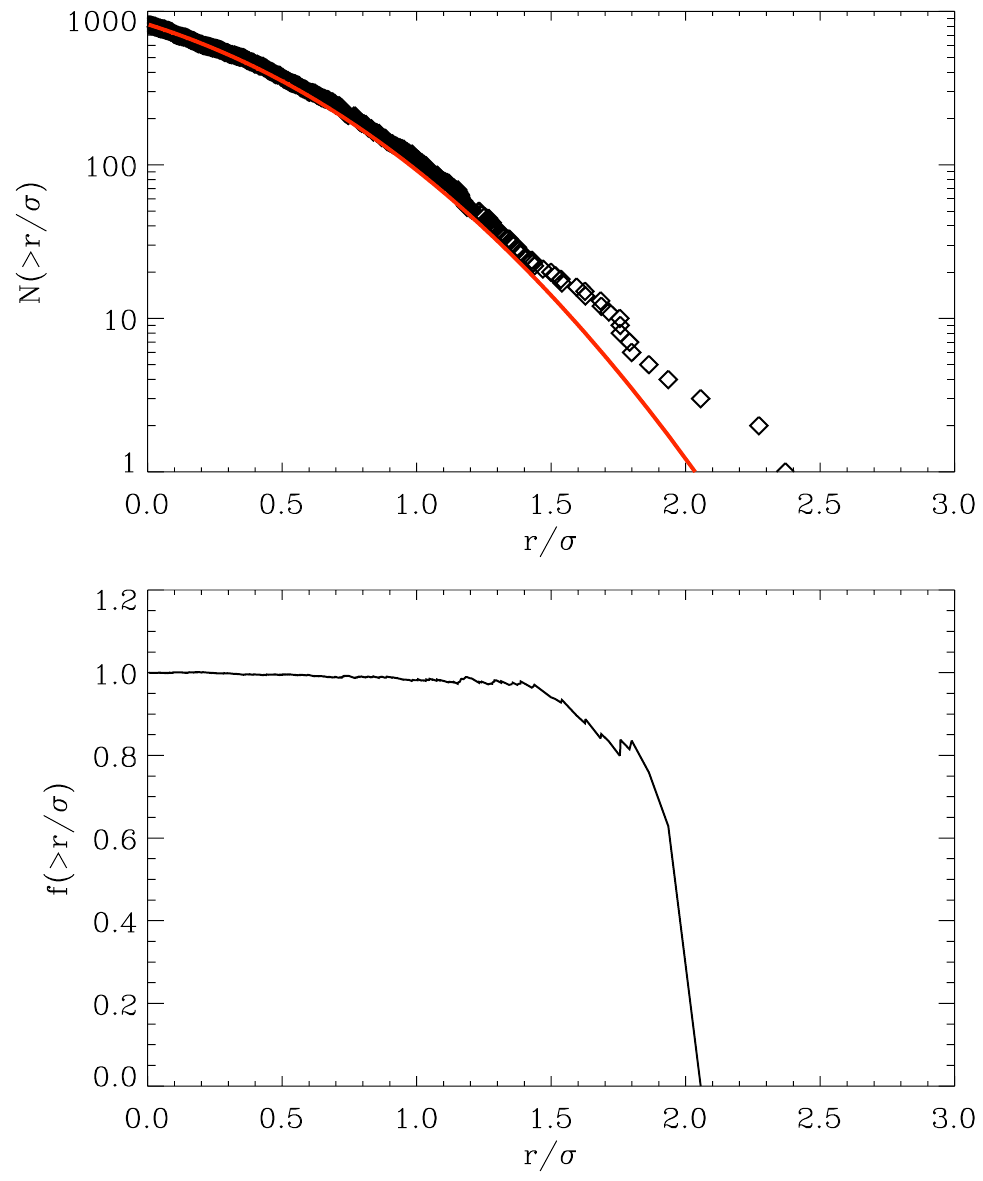}
   \caption{Upper panel: cumulative histograms of $| r_i / \sigma_i |$ for SARG 
data. The diamonds show the observed data, and the solid curve 
shows the result expected for a Gaussian-distributed noise. 
Lower panel: ratio $f$ of the observed to the expected histograms, 
indicating the fraction of “good” data points. An excess of outliers is evident for $| r_i / \sigma_i | \gtrsim 1.5$.
      }
   \label{fig:outliers}
   \end{figure}
%
       \begin{figure*}
   \centering
   \includegraphics[width=18cm]{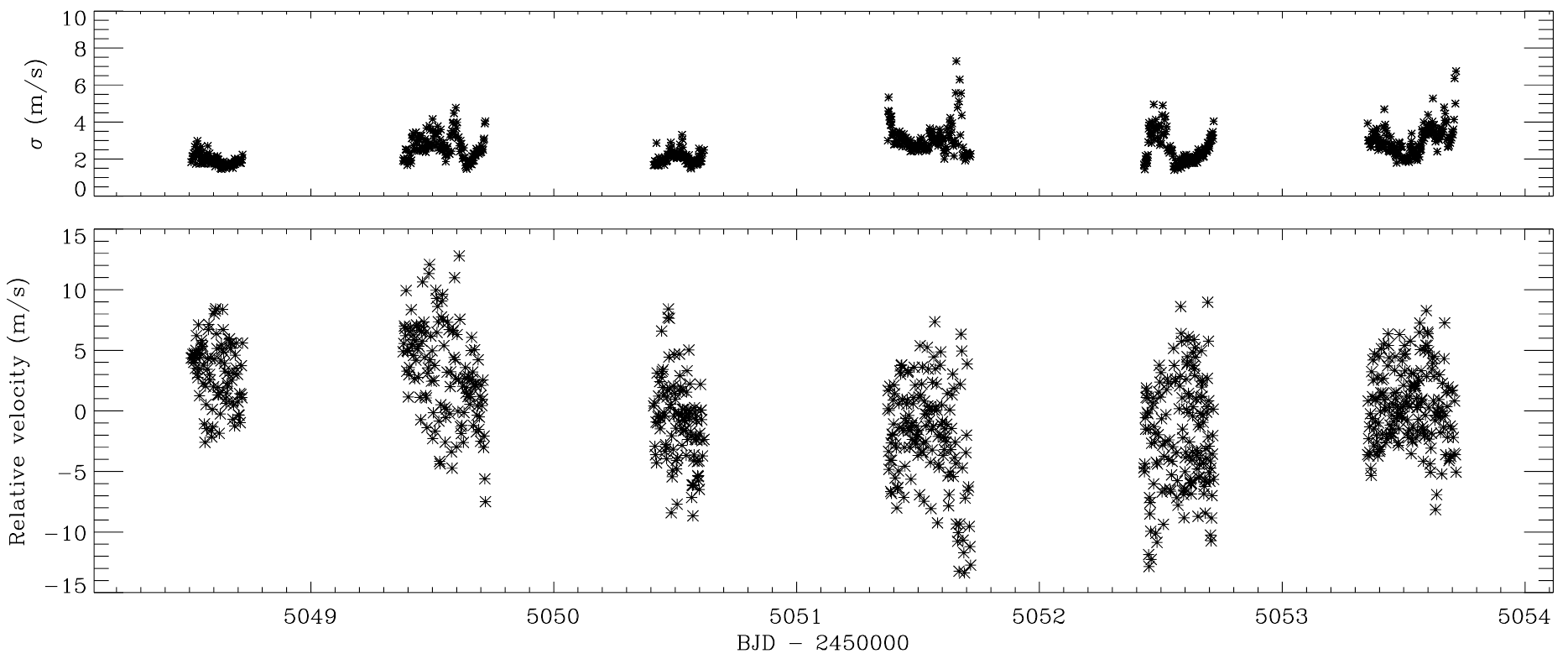}
      \caption{Radial velocity measurements of $\beta$ Aql for the entire time of observation, obtained with the iSONG code from the SARG spectra (lower panel) and their corresponding noise-scaled and outliers-corrected uncertainties $\sigma_i$ (upper panel).
              }
        \label{fig:total_ts}
   \end{figure*}
       \begin{figure*}
   \centering
   \includegraphics[width=18cm]{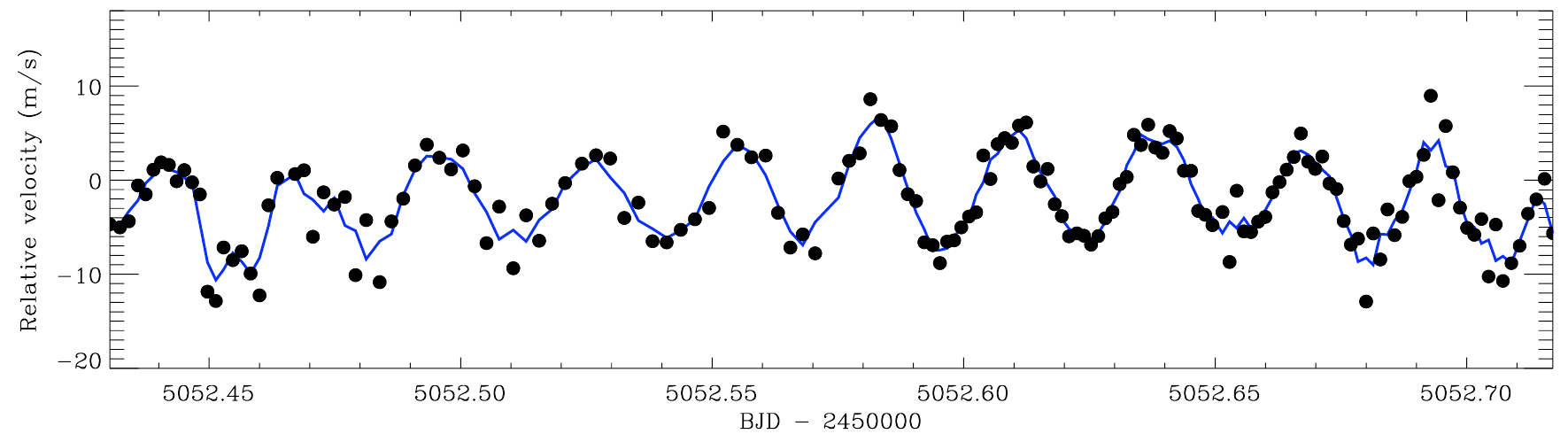}
      \caption{Detail of the fifth night of observation of radial velocity measurements obtained with the 
iSONG code from the SARG spectra (filled black circles). The solid blue line represents a 3.30 min wide smoothing to enhance the p-mode oscillations pattern.
              }
        \label{fig:5th_night}
   \end{figure*}
%
       \begin{figure*}
   \centering
   \includegraphics[width=18cm]{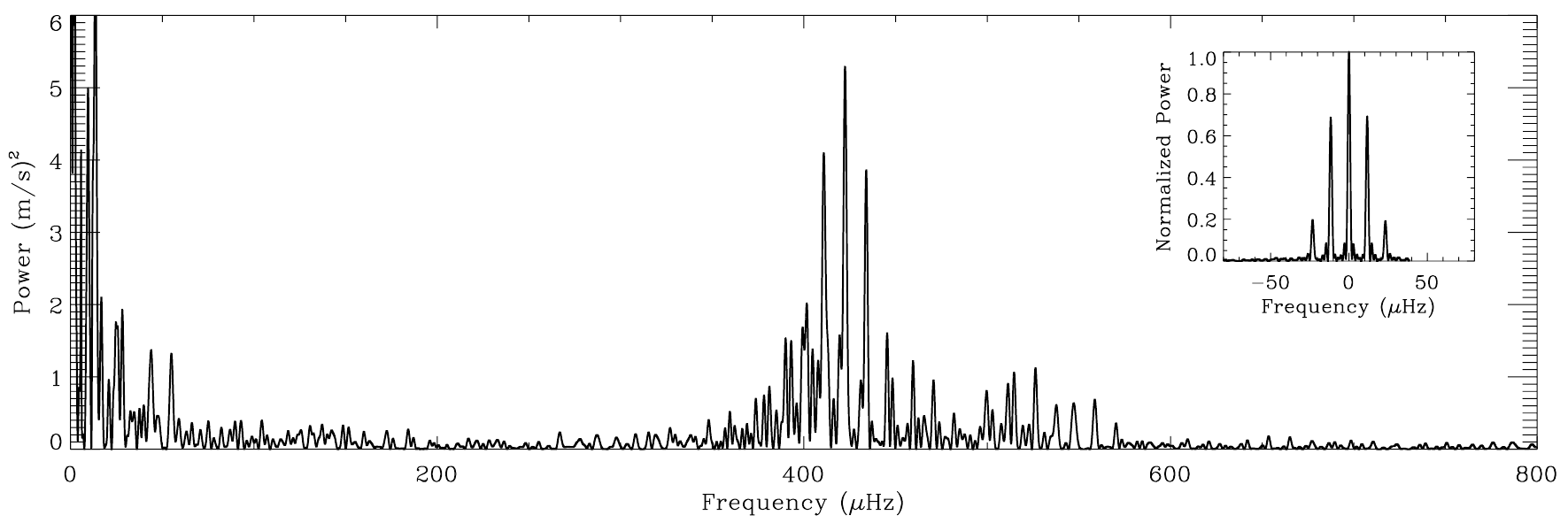}
      \caption{Power spectrum of the weighted radial velocity measurements of $\beta$ Aql extracted with the iSONG code from the SARG spectra. An excess of power is clearly visible, with a maximum centered at 422 $\mu$Hz. The inset shows the normalized power spectrum of the window function for a sine-wave signal of amplitude 1 m s$^{-1}$, sampled in the same way as the observations.}
        \label{fig:wps}
   \end{figure*}
This procedure, namely the pre-whitening, was carried out for the strongest peaks in the oscillation spectrum in the frequency range 0 - 1.5 mHz, until the spectral leakage into high frequencies from the remaining power was negligible \citep[see also][]{Leccia07,Bonanno08}. This left us with a time series of residual velocities, $r_i$, that reflects the noise properties of the measurements.  (ii) We then analyzed the ratio $r_i$/$\sigma_i$, expected to be Gaussian-distributed, so that the outliers correspond to the data points that exceed the given distribution. The cumulative histogram of the residuals is shown as a solid red curve in the upper panel of Fig.~\ref{fig:outliers}, indicating the fraction of “good” data points.
An excess of outliers is evident for $| r_i / \sigma_i | \gtrsim 1.5$. In this step we used the theoretical Gaussian function for the cumulative histogram, given by the expression
\begin{equation}
F (x_i) = \frac{N}{M} \left[ 1 - erf \left(x_i / x_0 \right) \right],
\end{equation}
where 
\begin{equation}
erf (x) = \frac{2}{\sqrt{\pi}} \int^{x}_0 {e^{t^2} dt}
\end{equation}
is the error function, $N$ is the total number of data points, $M = max \left[ 1 - erf \left( x_i / x_0 \right) \right]$, and $x_i = | r_i / \sigma_i |$. The zero point fixed to $x_0 = 0.89$ allows us to adjust the fit for a noise-optimized distribution, which means that the chosen distribution minimizes the noise level in the weighted PS \citep[see][for more details on different kinds of weight optimizations]{Bedding07hydri}. The reason for this optimization relies on the possibility to improve the mode identification by increasing the S/N of the frequency peaks. (iii) The lower panel of Fig.~\ref{fig:outliers} shows the ratio $f$ of the values of the observed points to the corresponding ones of the cumulative distribution function, i.e. the fraction of data points that could be considered as “good” observations, namely those that are close to unity. The quantities $w_i = 1/(\sigma^{2}_i f)$ were adopted as weights in the computation of the weighted PS. 

The time-series of the whole data set is presented in Fig.~\ref{fig:total_ts} (lower panel) with the corresponding noise-scaled and outliers-corrected uncertainties $\sigma_i$ (upper panel). Fig.~\ref{fig:5th_night} shows the details of the oscillation observations during the fifth night, overlaid with a solid blue curve representing a smoothing of 3.30 min for enhancing the $p$-mode oscillations pattern. The final data point number of the full observation was reduced with respect to the number of observed spectra owing to a consistent improvement of the oscillation envelope, so that a total of 818 data points was taken into account for computing the time-series analysis.

\section{Time-series analysis}
\label{sc:sec3}
The amplitude spectrum of the velocity time series 
was calculated as a weighted least-squares ﬁt of sinusoids \citep{Frandsen95,Arentoft98,Bedding04cen,Kjeldsen05cenB} with a weight assigned 
to each point according to its uncertainty estimate obtained from the radial velocity measurement, as explained in Section~\ref{sc:sec2-1}. The result is shown in Fig.~\ref{fig:wps}, where a clear excess of power around 420 $\mu$Hz is visible, with the typical pattern for p-mode oscillations in a G9.5 subgiant star. This feature is apparent in the power spectra of individual nights, and its frequency agrees with theoretical expectations, as we will discuss in Section~\ref{sec:theor}. To determine the S/N of the peaks in the PS, we measured the noise level $\sigma_{\mathrm{amp}}$ in the amplitude spectrum in the range 1200 - 1500 $\mu$Hz, far from the excess of power. By means of the new weights introduced above, it was reduced from $14.4$ cm s$^{-1}$ to the final value of $12.8$ cm~s$^{-1}$, which corresponds to a noise level in the PS of $0.02$~m$^{2}$~s$^{-2}$. Since this is based on $818$ measurements, we can deduce the velocity precision on the corresponding time-scales using the relation $\sigma_{\mathrm{RMS}} = \sigma_{\mathrm{amp}} \sqrt{N/\pi}$, as derived by \cite{Kjeldsen95}, which gives a scatter per measurement of $2.1$~m~s$^{-1}$.

However, particular care has to be taken with the noise evaluation within the region of solar-like oscillations. Indeed, by evaluating the noise in the amplitude spectrum in the intervals $100-300$ $\mu$Hz and $600-800$ $\mu$Hz, just below and above the excess of power, we obtained the two noise levels $\sigma^{100-300}_\mathrm{amp} = 26.2$ cm s$^{-1}$ and $\sigma^{600-800}_\mathrm{amp} = 15.6$ cm s$^{-1}$, which appear to be quite different. We then adopted a noise decaying accordingly to a linear trend law, ranging between the two values within the region $200 - 700$ $\mu$Hz.

\subsection{Search for a comb-like pattern}
The mode frequencies for low-degree, high radial order p-mode oscillations in Sun-like stars are reasonably well approximated by the asymptotic relation \citep{Tassoul80}
\begin{equation}
\nu \left( n, \ell \right) = \Delta\nu \left( n + \frac{\ell}{2} + \epsilon \right) - \ell(\ell+1) \delta\nu_{\mathrm{02}}/6 \, ,
\label{eq:asymp}
\end{equation}
where n and $\ell$ are integers that define the radial order and angular degree of the mode, respectively, $\Delta \nu =$ $\langle \nu_{\mathrm{n,\ell}} - \nu_{\mathrm{n-1,\ell}} \rangle$ is the so-called mean large frequency separation and reflects the average stellar density, $\delta\nu_{\mathrm{02}} =$ $\langle \nu_{\mathrm{n,0}} - \nu_{\mathrm{n-1,2}} \rangle$ is the small frequency separation, a quantity sensitive to the sound speed gradient near the core, and $\epsilon$ is a quantity on the order of unity sensitive to the stratification of the surface layers. 
On attempting to find the peaks in our power spectrum that match the asymptotic relation, we were severely hampered by the single-site window function, whose power is visible in normalized units in the inset of Fig.~\ref{fig:wps}, giving an effective observation time of $\sim 1.80$ days. As is well known, daily gaps in a time series produce aliases in the power spectrum at spacings $\pm 11.57$ $\mu$Hz and multiples, which are difficult to disentangle from the genuine peaks. Various methods to search for regular series of peaks have been discussed in the literature, such as autocorrelation, comb response and histograms of frequencies. To find a starting value for the $\Delta\nu$ investigation we used the comb-response function method, where a comb-response function $CR \left(\Delta\nu\right)$ is calculated for all sensible values of $\Delta\nu$ \citep[see][for details]{Kjeldsen95boo}, representing a generalization of the PS of a PS and consequently allowing us to search for any regularity in the spectral pattern. In particular we used the generalized comb-response function discussed in \cite{Bonanno08}
\begin{equation}
CR \left( \Delta\nu \right) = \prod^{N}_{n=1} \left[ PS\left( \nu_0 \pm \frac{2n -1}{2} \Delta\nu \right) PS \left( \nu_0 \pm n \Delta\nu \right) \right]^{\frac{1}{2^{n-1}}},
\end{equation}
so that a peak in the $CR$ at a particular value of $\Delta\nu$ indicates the presence of a regular series of peaks in the PS, centered at $\nu_{\mathrm{max}}$ and having a spacing of $\Delta\nu/2$. It differs from a correlation function because the product of individuals terms is considered rather than the sum. For actual calculations we used  $N = 2$ but we checked that our result was stable for  $N>2$ as well.

\begin{figure}
   \centering
   \includegraphics[width=9cm]{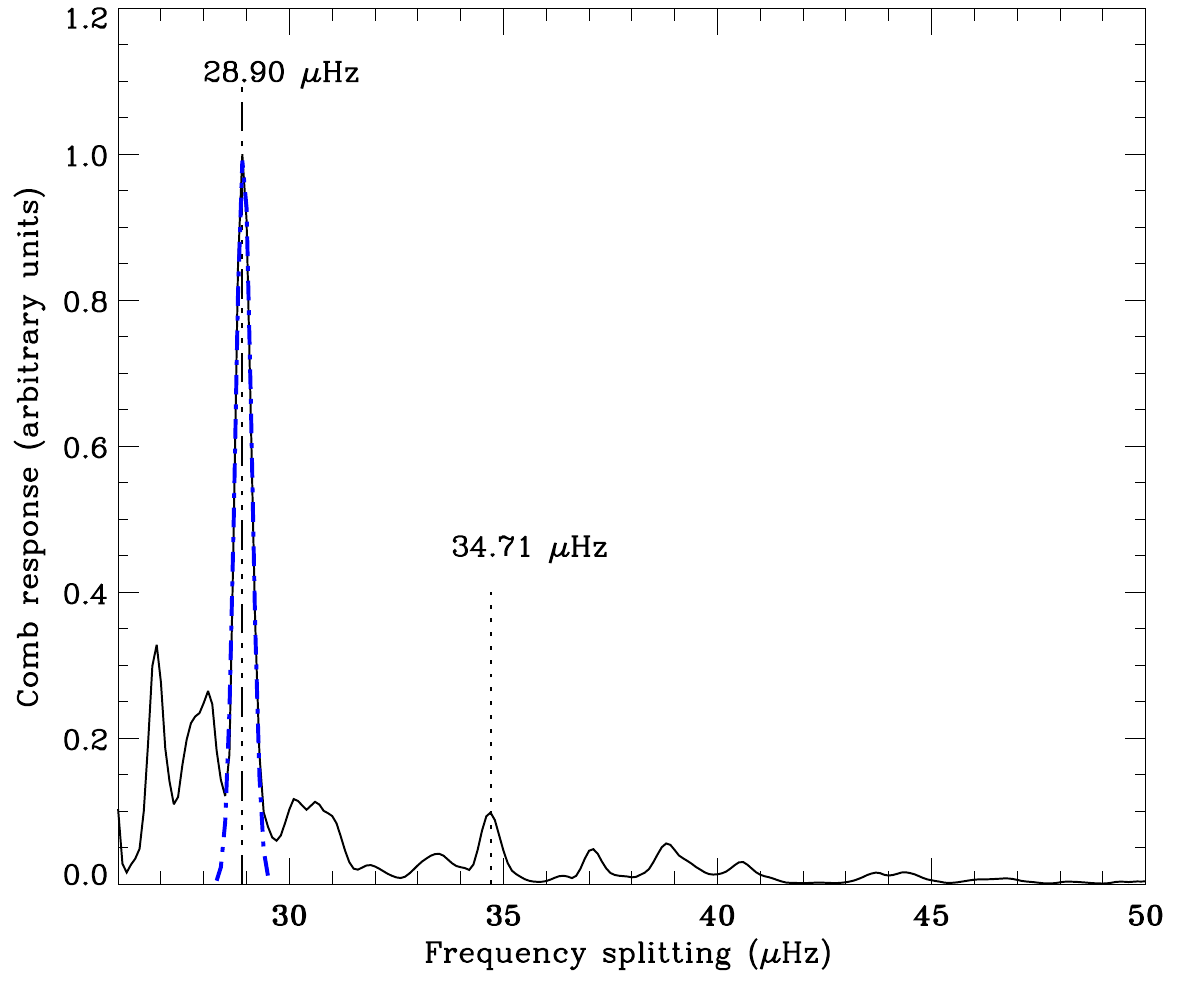}
      \caption{Cumulative comb-response obtained as the sum of 
the individual comb-responses for each central frequency $\nu_{0}$ with amplitude $> 0.5$ m s$^{-1}$ (S/N $\gtrsim 3.5$) in the amplitude spectrum. 
The maximum peak is centered at $\Delta\nu = 28.90 \pm 0.45$ $\mu$Hz where the uncertainty is evaluated as the FWHM of the Gaussian used to fit the peak (blue dot-dashed curve). The second marked peak on the right, centered at $34.71$ $\mu$Hz, represents three times the daily spacing.}
        \label{fig:comb}
   \end{figure}

To reduce the uncertainties caused by noise, only peaks above $300$ $\mu Hz$ and with amplitude $> 0.5$ m s$^{-1}$ in the amplitude spectrum, corresponding to a S/N $\gtrsim 3.5$, were used to compute the $CR$. We determined the local maxima of the response function $CR \left( \Delta\nu \right)$ in the range 8 $\leq \Delta\nu \leq$ 50 $\mu$Hz as first step. In this case the result was seriously affected by two very strong peaks at 11.57 $\mu$Hz and 23.14 $\mu$Hz, corresponding to once and twice the value of the daily gap, respectively, which consistently reduced the strength of the peak corresponding to $\Delta \nu$. We then decided to restrict the search range to 26 $\leq \Delta\nu \leq$ 50 $\mu$Hz as second step to completely exclude the daily alias peaks. The resulting cumulative comb-response function for this range, obtained by summing the contributions of all the response functions, had the most prominent peak centered at $28.90 \pm 0.45$ $\mu$Hz as shown in Fig.~\ref{fig:comb}, where the uncertainty was computed by considering the FWHM of the Gaussian used to fit the peak, and represented by the blue dot-dashed curve. The peak corresponding to three times the daily spacing is also visible in the right side of the plot, centered at $34.71$ $\mu$Hz.
\subsection{Oscillation frequencies}
\label{sec:of}
\begin{figure}
   \centering
   \includegraphics[width=9cm]{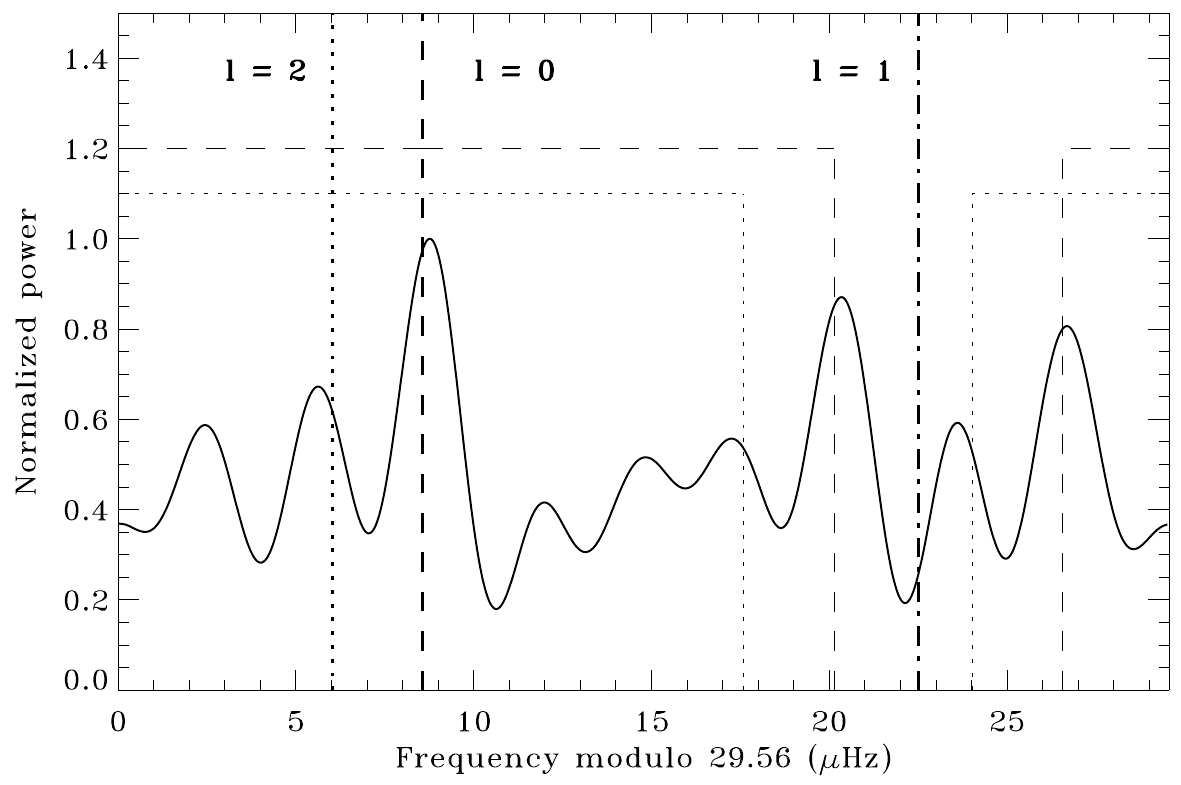}
   \caption{Folded PS in normalized units in the case of $\Delta\nu = 29.56$ $\mu$Hz. The overlaid ridges represent the result for $\ell = 0,1,2$ mode degrees as derived by a least-squares fit to the asymptotic relation, which are marked as dotted, dashed and dot-dashed lines, respectively. The daily side-lobes for $\pm 11.57$ in the case of $\ell = 0,2$ are also clearly visible and marked with the same line-style for each mode degree.
              }
   \label{fig:fold}
   \end{figure}

The comb-response function provided a guess of the large separation, which was then used as a starting point to investigate the most reliable value that could represent the observed data. The investigation involved a parallel checking and trade-off between two different methods, i.e. the folded PS and the échelle diagram. Firstly we computed the folded PS, namely the PS collapsed at the large separation value, for different values of $\Delta\nu$ subsequent to the guess number. The final result for the folding was computed for $\Delta\nu = 29.56$ $\mu$Hz and is shown in Fig.~\ref{fig:fold}, where the peaks corresponding to $\ell = 0$ and $\ell = 2$ ridges are marked by a dashed and a dotted line, respectively. The daily side-lobes for $\pm 11.57$ $\mu$Hz are also clearly visible and are marked with the same line-styles. The overlaid ridges represent the result of a least-squares fit to the asymptotic relation~(\ref{eq:asymp}). It is noticeable how the $\ell = 1$ ridge, marked by a dot-dashed line, does not correspond to any peak in the folded PS. This could be caused by avoided crossings, although it could also be explained by a wrong identification of the modes, as we will discuss in more detail in Section~\ref{sec:theor}. We note that the $\ell = 2$ ridge appears to be slightly shifted with respect to the position of the maximum of the relative peak, a result that can be explained by the limit of our frequency resolution.
%

    \begin{figure}
   \centering
   \includegraphics[width=9cm]{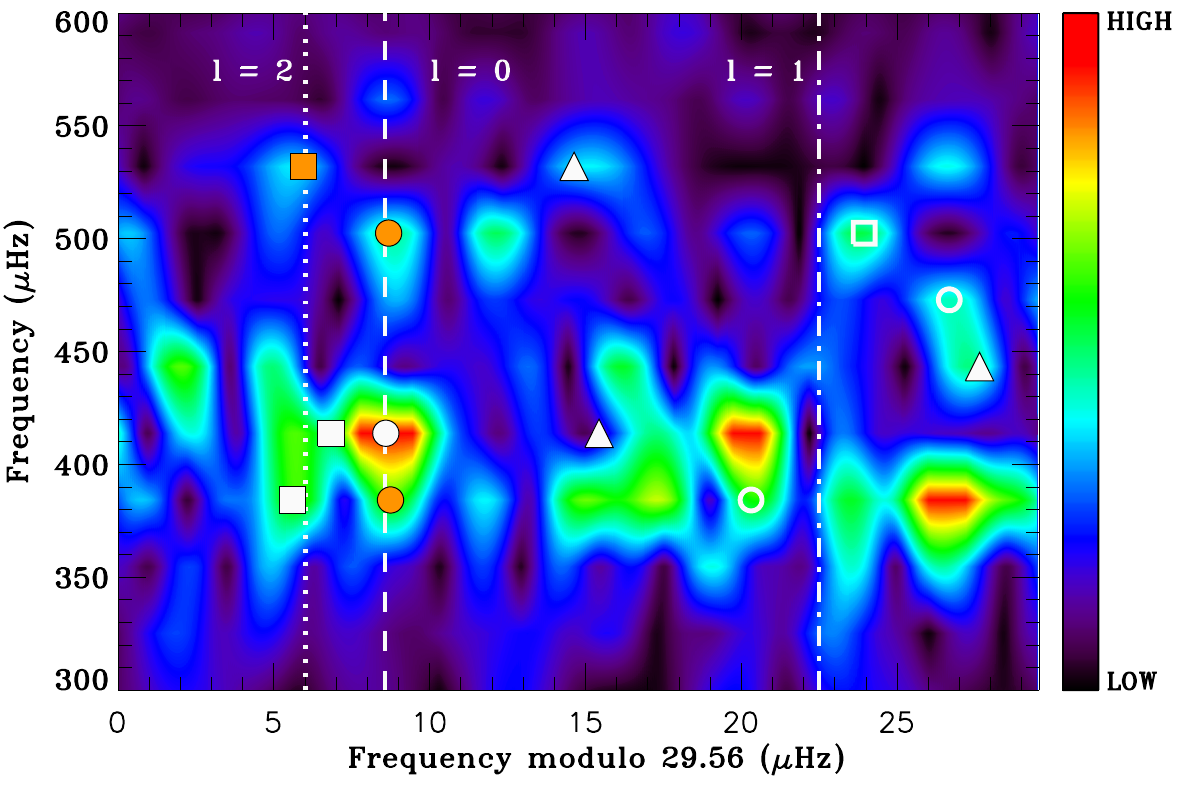}
   \caption{\'{E}chelle diagram overlaid on the amplitude spectrum with a colored background scale. The filled symbols (white and orange) represent the identified modes for $\ell = 0$ (circles), $\ell = 1$ (triangles) and $\ell = 2$ (squares). The orange symbols are the frequencies shifted for the daily gap of $\pm 11.57$ $\mu$Hz while the open symbols correspond to the original unshifted values. The ridges derived from the fit to the asymptotic relation~(\ref{eq:asymp}) are also marked.
              }
   \label{fig:ech}
   \end{figure}
%
The second method adopted for the investigation of the large separation is represented by the well known échelle diagram, an essential tool for frequency identification in asteroseismic data. The échelle diagram was computed for a list of 9 high S/N ($\gtrsim 3.5$) frequencies directly obtained from the PS by means of the CLEAN algorithm, although a further consideration regarding this aspect is required to explain the way the final list was attained. In fact, two different ways of CLEANing the PS were adopted for this work. In the first case the algorithm was applied to the weighted PS for a sufficiently large number of frequencies ($\simeq$ 50 to find all the peaks with S/N $> 3$) in the range 0.01 - 700 $\mu$Hz, then the resulting values were restricted to the interval 200 - 700 $\mu$Hz, obtaining a list of 20 frequencies. For the second case the PS region below 200 $\mu$Hz was at first completely pre-whitened. Then 20 frequencies were identified on the new resulting weighted PS in the range 200 - 700 $\mu$Hz. Comparing both lists of frequencies, the most unstable ones, i.e. those that were not present in both lists, were rejected and only the first nine frequencies were considered. These frequencies, showing an amplitude $> 0.5$ m s$^{-1}$ (or S/N $\gtrsim 3.5$), were almost the same in both cases, however, the first frequency list was selected because the chance of the identification of one more $\ell = 2$ mode was possible for this case.
In Fig.~\ref{fig:ech} we show the final result for the échelle diagram superimposed on a colored scale background representing the amplitude spectrum, where the plot was computed in the same way as in \cite{Mathur11}. The filled symbols (white and orange) represent the 10 identified modes for $\ell = 0$ (circles), $\ell = 1$ (triangles) and $\ell = 2$ (squares); the orange symbols are the corrected frequencies, i.e. those shifted for the daily gap of $\pm 11.57$ $\mu$Hz from their original values (the corresponding open symbols) and reported in Table~\ref{table:1}. The aliasing considerably affects the amplitude spectrum with the presence of several fictitious peaks, which appear as strong spots in the diagram without frequency symbols overlaid. In particular, the two daily side-lobes arising from the strongest mode ($\ell = 0$) are clearly visible as two red spots on the right-hand part of the diagram.
The uncertainties on frequencies are listed in Table~\ref{table:1} and were evaluated by using the analytical relation provided by \cite{Montgomery99}
\begin{equation}
\sigma (\nu) = \sqrt{\frac{6}{N}} \frac{1}{\pi T} \frac{\langle \sigma_\mathrm{v} \rangle}{A} \simeq \frac{0.16}{A_\mathrm{m / s}} \,\, \mu \mbox{Hz} \, ,
\label{eq:sig_frq}
\end{equation}
where $N = 818$ is the total number of data points, $T = 5.21$ is the total duration of the run in days, $\langle \sigma_\mathrm{v} \rangle  = 2.72$ m s$^{-1}$ is the average uncertainty on radial velocity each data point and A is the amplitude per mode as derived in Section~\ref{sec:modamp}. This relation holds exactly for coherent oscillations, hence we remark that the estimated uncertainties represent only a lower limit to the real uncertainty value because these oscillations are not fully coherent. An upper limit to these uncertainties can be fixed to the formal frequency resolution, given as the reciprocal of the total duration of the run, which is $2.2$ $\mu$Hz for this data set. The complete identification of the p modes is reported in Table~\ref{table:1} together with their S/N, where only three frequencies out of nine were shifted by the daily alias. The $\ell = 1$ modes reported without any radial order number are potential candidates for mixed modes.
\begin{table}
\caption{Mode identification for $\beta$ Aql, in the frequency range 300-600 $\mu$Hz. The modes listed show an amplitude $> 0.5$ m s$^{-1}$ (or S/N $\gtrsim 3.5$). The corrected frequencies reported in the fourth column include the frequencies shifted for the daily gap of $\pm 11.57$ $\mu$Hz. The fifth column represents the uncertainties as derived by means of the analytical relation~(\ref{eq:sig_frq}). The $\ell = 1$ frequencies reported without any radial order are possible avoided crossings.}             
\label{table:1}    
\centering                         
\begin{tabular}{c c r l c}       
\hline\hline                
$\ell$ & n & S/N & Corrected frequency$^{\mathrm{a}}$ ($\mu$Hz) & Uncertainty$^{\mathrm{b}}$ ($\mu$Hz)\\    
\hline                        
   0 & 12 & 6.4 & 393.02 (+11.57) & 0.22\\     
   0 & 13 & 11.3 & 422.43 & 0.22\\
   0 & 16 & 3.7 & 511.20 (-11.57) & 0.34\\
\hline
   1 & - & 4.0 & 429.27 & 0.16\\ 
   1 & - & 4.7 & 471.04 & 0.20\\ 
   1 & - & 3.5 & 546.71 & 0.38\\ 
\hline
   2 & 11 & 3.5 & 389.87 & 0.22\\ 
   2 & 12 & 3.5 & 420.67 & 0.21\\ 
   2 & 16 & 5.1 & 538.03 (-11.57) & 0.42\\ 
\hline                                
\end{tabular}
\begin{list}{}{}
\item[$^{\mathrm{a}}$] The raw frequencies can be evaluated by adding the daily frequency reported in parentheses.
\item[$^{\mathrm{b}}$] The actual uncertainties can be several times larger up to the limit of the formal resolution of $2.2$ $\mu$Hz, because the modes are not coherent.
\end{list}
\end{table}
\begin{table}
\caption{Asymptotic parameters as derived by a linear weighted least-squares fit to the asymptotic relation~(\ref{eq:asymp}).}             
\label{table:2}
\centering                       
\begin{tabular}{c c c }   
\hline\hline               
$\Delta\nu$ ($\mu$Hz) & $\delta\nu_{\mathrm{02}}$ ($\mu$Hz) & $\epsilon$ \\  
\hline                      
   $29.56 \pm 0.10$ & $2.55 \pm 0.71^{a}$ & $1.29 \pm 0.04$ \\ 
\hline                      
\end{tabular}
\begin{list}{}{}
\item[$^{\mathrm{a}}$] The most likely value for the small separation was derived by using the definition from the asymptotic relation with the couples of frequencies $(\nu_{\mathrm{12,0}},\nu_{\mathrm{11,2}})$, $(\nu_{\mathrm{13,0}},\nu_{\mathrm{12,2}})$ and $(\nu_{\mathrm{16,0}},\nu_{\mathrm{16,2}} - \Delta \nu)$.
\end{list}
\end{table}

By means of a linear weighted least-squares fit to the asymptotic relation of the $\ell = 0$ frequencies, the final value of $\Delta\nu = 29.56 \pm 0.10$ $\mu$Hz was obtained, together with the constant $\epsilon = 1.29 \pm 0.04$. The most likely value for the small separation was derived by using the definition from the asymptotic relation with the frequency pairs $\left(\nu_{\mathrm{12,0}},\nu_{\mathrm{11,2}}\right)$, $\left(\nu_{\mathrm{13,0}},\nu_{\mathrm{12,2}}\right)$ and $\left(\nu_{\mathrm{16,0}},\nu_{\mathrm{15,2}}\right)$, where $\nu_{\mathrm{15,2}} = 508.47$ $\mu$Hz was computed directly from $\nu_{\mathrm{16,2}} = 538.03$ $\mu$Hz by adopting our value of the large separation. This led to $\delta\nu_{\mathrm{02}} = 2.55 \pm 0.71$ $\mu$Hz, as reported in Table~\ref{table:2}, but because it is comparable to the frequency resolution, its uncertainty is relatively high (Fig.~\ref{fig:ech}). As a consequence, the reliability of this result requires additional investigations and a longer data set. 

    \begin{figure}
   \centering
   \includegraphics[width=9cm]{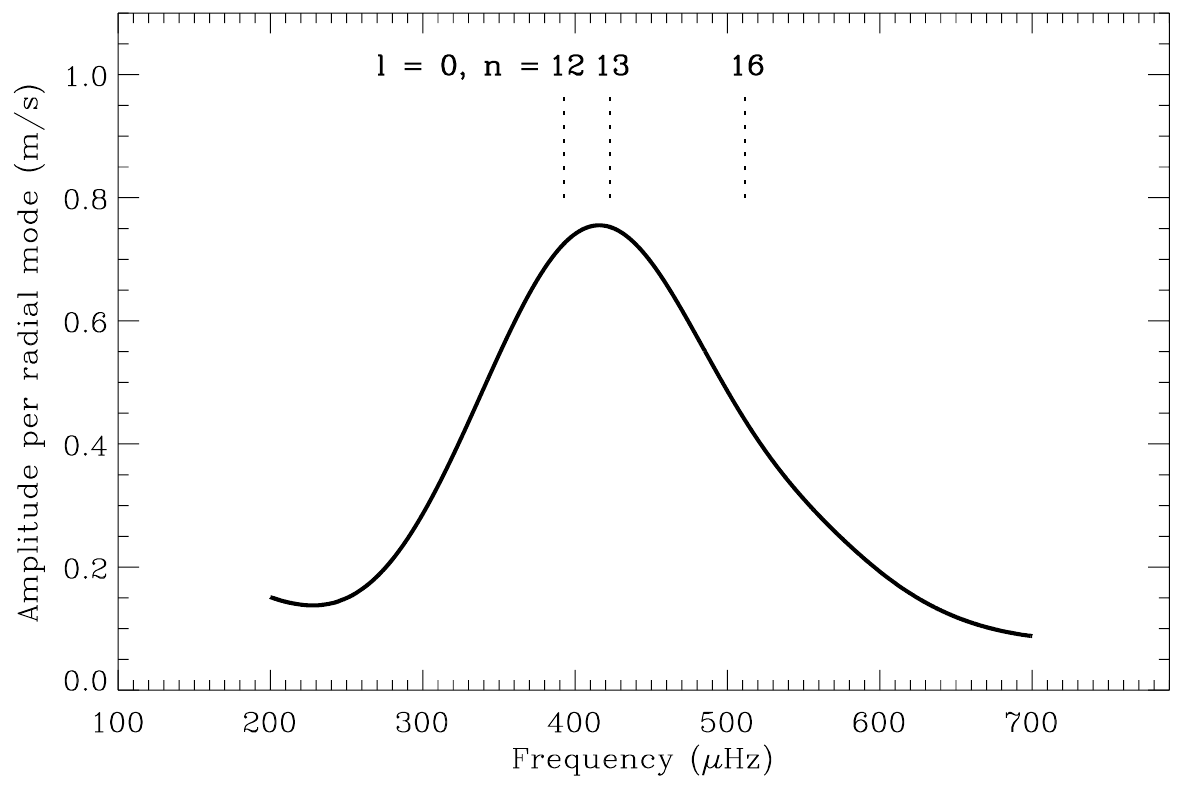}
   \caption{Smoothed amplitude spectrum showing the amplitude per radial mode computed in the range 200 - 700 $\mu$Hz. The maximum amplitude $A_{\mathrm{max}} = 76 \pm 13$ cm s$^{-1}$ occurs at $\nu_{\mathrm{max}} = 416$ $\mu$Hz. The positions of the identified $\ell = 0$ frequencies as derived from asymptotic relationship~(\ref{eq:asymp}) are also marked.
              }
   \label{fig:amp}
   \end{figure}
%
\subsection{Mode amplitudes}
\label{sec:modamp}
The evaluation of mode amplitudes is interesting for the discussion on the L/M scaling by extrapolating from the Sun, as we will discuss in Sec.~\ref{sec:theor}. 
As is known from the theory of solar-like oscillations, the amplitudes of individual modes are affected by the stochastic nature of the excitation and damping \citep[e.g. see][]{Kjeldsen95}. To measure the oscillation amplitude per mode in a way that is independent of these effects, we followed the approach explained in  \cite{Kjeldsen05cenB,Kjeldsen08baql}. This involves the following steps: (i) heavy smoothing of the weighted PS by convolving it with a Gaussian whose FWHM is fixed to $4 \Delta\nu$ (which is the value considered as a standard that allows comparisons since the amount of smoothing affects the exact height of the smoothed amplitude spectrum), which is large enough to produce a single hump of power that is insensitive to the discrete nature of the oscillation spectrum; (ii) conversion of the smoothed PS to power density spectrum (PSD) by multiplying by the effective length of the observing run (that is given by the reciprocal of the area under the spectral window in power, namely 6.40 $\mu$Hz for this data set); (iii) subtraction of the background noise, which we computed as a linear trend in the interval $200-700$ $\mu$Hz, ranging from $26.2$ cm s$^{-1}$ to $15.6$ cm s$^{-1}$; (iv) multiplication of the result by $\Delta\nu / c$ where $c = 4.09$ (which represents the effective number of modes that fall in each segment of length $\Delta\nu$ as evaluated in the case of radial velocities) and taking the square root to convert to amplitude per oscillations mode. The result is shown in Fig.~\ref{fig:amp} for the range $200-700$ $\mu$Hz, where $A_{\mathrm{max}} = 76 \pm 13$ cm s$^{-1}$ centered at $\nu_{\mathrm{max}} = 416$ $\mu$Hz, which is assumed to be the frequency of maximum power and agrees with the result of \cite{Kjeldsen08baql}. The uncertainty on the amplitude is evaluated by means of the analytical estimation relation
\begin{equation}
\sigma(A) = \sqrt{\frac{2}{N}} \hspace{2px} \langle \sigma_\mathrm{v} \rangle
\label{eq:sig_amp}
\end{equation}
\citep{Montgomery99}, where, as for the frequencies case, $N = 818$ is the total number of data points and $\langle \sigma_\mathrm{v} \rangle$ is once more the average uncertainty on each data point. Again, the uncertainty in amplitude represents a lower limit to the real value. The amplitude distribution is evaluated for the radial modes only, but the calculation of the amplitude in the case of $\ell = 1$ and $\ell = 2$ modes is straightforward, namely it can be obtained by multiplying the result for a factor of 1.35 and 1.02 respectively, representing the relative strength given by the spatial response function \citep[see][for more details]{Kjeldsen08baql}. The result derived in this work is not far from the value obtained by \cite{Kjeldsen08baql}.

\section{Discussion}
\label{sec:theor}
A complete discussion on the evolutionary state of this star goes beyond the scope of this work. Nevertheless, the identified modes provided a reliable estimate for the mean large separation, as described in Section~\ref{sec:of}, which can be used to derive the scaled mass for this star according to the fact that $\Delta\nu$ scales approximately with the square root of the mean density of the star \citep{Cox81}. From the scaling relation extrapolating from the solar case \citep[e.g. see][]{Bedding07hydri}
\begin{equation}
\Delta\nu = \left( \frac{M}{M_\odot} \right)^{\frac{1}{2}} \left( \frac{R}{R_\odot} \right)^{-\frac{3}{2}} \Delta\nu_\odot \, ,
\label{eq:scale} 
\end{equation}
where the radius is provided by \cite{Nordgren1999} and $\Delta\nu_\odot = 134.9$ $\mu$Hz, we obtained a mass of $M = 1.36 \pm 0.17$ $M_\odot$, which agrees very well with the value found by \cite{Fuhrmann04}. By considering a luminosity of $L = 5.63 \pm 0.16$ $L_\odot$, as derived by means of visual magnitude \citep{Oja86}, bolometric correction \citep{Flower96}, Sun bolometric magnitude \citep{Cox2000} and \textit{Hipparcos} parallax \cite[]{HIPPARCOS}, and the scaled value for the mass, we were able to compute the value for the amplitude by using the relation provided by \cite{Kjeldsen95}:
\begin{equation}
A_{\mathrm{osc}} = \frac{L/L_\odot}{M/M_\odot} (23.4 \pm 1.4) \hspace{3px} \mbox{cm s$^{-1}$},
\end{equation}
which gives $A_{\mathrm{osc}} =$ $97 \pm 6$ cm s$^{-1}$, which fairly agrees with the value obtained in Section~\ref{sec:modamp}.
Moreover, the expected frequency of maximum power can be evaluated from scaling the acoustic cutoff frequency from the solar case. We computed this frequency by considering the relationship
\begin{equation}
\nu_{\mathrm{max}} = \frac{M/M_\odot}{\left( R/R_\odot \right)^3 \sqrt{T_{\mathrm{eff}} / 5777 K}} \hspace{2px} 3.05 \cdot 10^3 \mbox{$\mu$Hz}
\end{equation}
\citep{Kjeldsen95}, whose result gives $\nu_{\mathrm{max}} = 472 \pm 72$ $\mu$Hz, consistent with the value obtained on the smoothed amplitude spectrum.
\begin{table}
\caption{Global list of stellar parameters for $\beta$ Aql. }          
\label{table:3}     
\centering                      
\begin{tabular}{l l c}     
\hline\hline                
Stellar parameter & \multicolumn{1}{c}{Value} & Source\\
\hline
M & $1.36 \pm 0.17$ $M_\odot$ & This work$^{\mathrm{a}}$\\  
R & $3.05 \pm 0.13$ $R_\odot$ & \citep{Nordgren1999} \\ 
$\langle \rho \rangle$ & $0.0676 \pm 0.0004$ g cm$^{-3}$ & This work\\
Z & $0.015 \pm 0.002$ & SEEK\\ 
X & $0.70 \pm 0.02$ & SEEK\\ 
Age & $2.43^{+3.56}_{-0.32}$ Gyr & SEEK\\ 
L & $5.63 \pm 0.16$ $L_\odot$ & This work\\ 
log g & $3.61^{+0.01}_{-0.02}$ & SEEK\\ 
$T_{\mathrm{eff}}$ & $5160 \pm 100$ K & \citep{Luck05}\\
$\Delta\nu$ & $29.56 \pm 0.10$ $\mu$Hz & This work\\
$\delta\nu_{\mathrm{02}}$ & $2.55 \pm 0.71$ $\mu$Hz & This work\\
$A_{\mathrm{max}}$ & $76 \pm 13$ cm s$^{-1}$ & This work\\
\hline
\end{tabular}
\begin{list}{}{}
\item[$^{\mathrm{a}}$] The corresponding parameter was derived directly in this work.
\end{list}
\end{table} 
Concerning the values for the mean large separation and the frequency of maximum power it is outstanding that they fit the power law relation of \cite{Stello09} quite well, which represents a considerable validation to the reliability of our results. Moreover, an independent measure for the large separation by using the Eq.~(\ref{eq:scale}), with the radius from \cite{Nordgren1999} and the mass from \cite{Fuhrmann04} is also compatible with the value presented in Section~\ref{sec:of}.

Lastly, an important note regards the dipole frequencies reported in this work, which are expected to deviate strongly from their asymptotic values, especially for an evolved subgiant star like $\beta$ Aql. Indeed, the presumed $\ell = 1$ modes identified here are possibly bumped because of avoided crossings. Although the lack of a clear $\ell = 1$ ridge in Fig.~\ref{fig:fold} could also be explained by a wrong identification of the modes, we are confident that the large separation is correct. The reason of our belief is that different tools for its investigation, such as the comb-response function, the folded PS, the échelle diagram and the fit to the asymptotic relation, show consistent results. In addition, its compatibility with the independent estimation and the agreement with the $\Delta \nu - \nu_\mathrm{max}$ power law relation mentioned in the above paragraph ensure a robust derivation of the result presented.
Nonetheless, we are talking about a very difficult star, such as $\nu$ Ind \citep{Carrier07}, because in the region of the HR diagram to which the star belongs, avoided crossings considerably hamper the p-mode identification \citep[see][for a summary on ground-based observations across the HR diagram]{Bedding2011}. Therefore, a theoretical confirmation is required before adopting the $\ell = 1$ frequencies reported in Table~\ref{table:1} as real frequencies of oscillations for mixed modes, and more observations by a multi-site ground-based project such as SONG are required to firmly solve the mode identification.

\subsection{Stellar parameters}
The SEEK package \citep{Quirion10} is developed for the analysis of asteroseismic data from the \textit{Kepler} mission and is able to estimate stellar parameters in a form that is statistically well-defined. It is based on a large grid of stellar models computed with the Aarhus Stellar Evolution Code (ASTEC), which allow us to derive additional stellar parameters giving as input astrophysical quantities such as $T_{\mathrm{eff}}$ \citep{Luck05}, $\log g$ \citep{Valenti05} and $[Fe/H]$ \citep{Fuhrmann04}, and the asteroseismic values derived in this work, i.e. the large and small separation and the frequency of maximum power. The output list of parameters for $\beta$ Aql is shown in Table~\ref{table:3}, where the 1-$\sigma$ error bars on the SEEK values are computed using the Bayesian evaluation of the posterior distributions. The mass, radius and luminosity computed by SEEK also agree with the values presented in this work.

\section{Conclusions}
\label{sec:conc} 
Our observations of $\beta$ Aql show an evident excess of power in the PS region centered at 415 $\mu$Hz, clearly very well separated from the low-frequency power, and with a position and amplitude that agree with expectations. Although consistently hampered by the single-site window, the comb analysis and the échelle diagram show clear evidence for regularity in the peaks at the spacing expected from the asymptotic theory. The complete identification of six high S/N modes for $\ell = 0,2$ led to a well-constrained mean large separation of $\Delta\nu = 29.56 \pm 0.10$~$\mu$Hz, compatible with a scaled value from the Sun and the value obtained by the power law relation \citep{Stello09}, and to a most likely value for the small separation of $\delta\nu_{\mathrm{02}} = 2.55 \pm 0.71$ $\mu$Hz, whose reliability has yet to be confirmed. 
The $\ell = 1$ modes found are presumably mixed modes but a theoretical confirmation is needed before adopting these values as real modes of oscillations for this star. Moreover, our results provide a valuable proof that oscillations in an evolved subgiant solar-like star show amplitudes that scale as $L/M$ by extrapolating from the Sun. 

This campaign of observations attained with SARG led to high-precision RV measurements by means of the iSONG code, which was used for the first time in this work. The time-series analysis of the given data set was able to provide for the first time global asteroseismic parameters and individual p modes together with the evidence for mixed modes. Moreover, this result will be extremely important to develop theoretical models for this star. Multi-site observation campaign with the SONG project is highly desirable in a near future. That would then allow us to explore the solar-like oscillations for this target in a detailed way by providing a large number of identified modes. The asteroseismic and astrophysical parameters of this star will then be constrained properly, yielding a deeper understanding of solar-like oscillations in the very difficult region of the HR diagram to which $\beta$ Aql belongs.

\begin{acknowledgements}
      We thank the Italian Foundation Galileo for the chance to acquire spectra at the TNG Italian Telescope in La Palma; the Department of Physics and Astronomy of the Aarhus University for hosting part of this work; Antonio Frasca for suggestions in the calculation of the stellar luminosity; Christoffer Karoff for the evaluation of additional stellar parameters with the SEEK package; the Sydney Institute for Astronomy (SIfA), School of Physics of the University of Sydney, for hosting part of this work; Tim Bedding and Dennis Stello for important discussions and advice in improving the paper.
\end{acknowledgements}

\bibliographystyle{aa} 
\bibliography{biblio} 

\begin{thebibliography}{46}
\expandafter\ifx\csname natexlab\endcsname\relax\def\natexlab#1{#1}\fi

\bibitem[{{Arentoft} {et~al.}(1998){Arentoft}, {Kjeldsen}, {Nuspl}, {Bedding},
  {Fronto}, {Viskum}, {Frandsen}, \& {Belmonte}}]{Arentoft98}
{Arentoft}, T., {Kjeldsen}, H., {Nuspl}, J., {et~al.} 1998, \aap, 338, 909

\bibitem[{{Baglin} {et~al.}(2006){Baglin}, {Michel}, {Auvergne}, \& {The COROT
  Team}}]{CoRoT06}
{Baglin}, A., {Michel}, E., {Auvergne}, M., \& {The COROT Team}. 2006, in ESA
  Special Publication, Vol. 624, Proceedings of SOHO 18/GONG 2006/HELAS I,
  Beyond the spherical Sun

\bibitem[{{Bedding} \& {Kjeldsen}(2006)}]{Bedding06}
{Bedding}, T. \& {Kjeldsen}, H. 2006, in ESA Special Publication, Vol. 624,
  Proceedings of SOHO 18/GONG 2006/HELAS I, Beyond the spherical Sun

\bibitem[{{Bedding}(2011)}]{Bedding2011}
{Bedding}, T.~R. 2011, ArXiv e-prints (arXiv:1107.1723v1)

\bibitem[{{Bedding} \& {Kjeldsen}(2003)}]{Bedding03}
{Bedding}, T.~R. \& {Kjeldsen}, H. 2003, \pasa, 20, 203

\bibitem[{{Bedding} \& {Kjeldsen}(2008)}]{Bedding08}
{Bedding}, T.~R. \& {Kjeldsen}, H. 2008, in Astronomical Society of the Pacific
  Conference Series, Vol. 384, 14th Cambridge Workshop on Cool Stars, Stellar
  Systems, and the Sun, ed. {G.~van Belle}, 21

\bibitem[{{Bedding} {et~al.}(2007){Bedding}, {Kjeldsen}, {Arentoft}, {Bouchy},
  {Brandbyge}, {Brewer}, {Butler}, {Christensen-Dalsgaard}, {Dall}, {Frandsen},
  {Karoff}, {Kiss}, {Monteiro}, {Pijpers}, {Teixeira}, {Tinney}, {Baldry},
  {Carrier}, \& {O'Toole}}]{Bedding07hydri}
{Bedding}, T.~R., {Kjeldsen}, H., {Arentoft}, T., {et~al.} 2007, \apj, 663,
  1315

\bibitem[{{Bedding} {et~al.}(2004){Bedding}, {Kjeldsen}, {Butler}, {McCarthy},
  {Marcy}, {O'Toole}, {Tinney}, \& {Wright}}]{Bedding04cen}
{Bedding}, T.~R., {Kjeldsen}, H., {Butler}, R.~P., {et~al.} 2004, \apj, 614,
  380

\bibitem[{{Bonanno} {et~al.}(2008){Bonanno}, {Benatti}, {Claudi}, {Desidera},
  {Gratton}, {Leccia}, \& {Patern{\`o}}}]{Bonanno08}
{Bonanno}, A., {Benatti}, S., {Claudi}, R., {et~al.} 2008, \apj, 676, 1248

\bibitem[{{Borucki} {et~al.}(2010){Borucki}, {Koch}, {Basri}, {Batalha},
  {Brown}, {Caldwell}, {Caldwell}, {Christensen-Dalsgaard}, {Cochran},
  {DeVore}, {Dunham}, {Dupree}, {Gautier}, {Geary}, {Gilliland}, {Gould},
  {Howell}, {Jenkins}, {Kondo}, {Latham}, {Marcy}, {Meibom}, {Kjeldsen},
  {Lissauer}, {Monet}, {Morrison}, {Sasselov}, {Tarter}, {Boss}, {Brownlee},
  {Owen}, {Buzasi}, {Charbonneau}, {Doyle}, {Fortney}, {Ford}, {Holman},
  {Seager}, {Steffen}, {Welsh}, {Rowe}, {Anderson}, {Buchhave}, {Ciardi},
  {Walkowicz}, {Sherry}, {Horch}, {Isaacson}, {Everett}, {Fischer}, {Torres},
  {Johnson}, {Endl}, {MacQueen}, {Bryson}, {Dotson}, {Haas}, {Kolodziejczak},
  {Van Cleve}, {Chandrasekaran}, {Twicken}, {Quintana}, {Clarke}, {Allen},
  {Li}, {Wu}, {Tenenbaum}, {Verner}, {Bruhweiler}, {Barnes}, \&
  {Prsa}}]{Borucki10}
{Borucki}, W.~J., {Koch}, D., {Basri}, G., {et~al.} 2010, Science, 327, 977

\bibitem[{{Butler} {et~al.}(2004){Butler}, {Bedding}, {Kjeldsen}, {McCarthy},
  {O'Toole}, {Tinney}, {Marcy}, \& {Wright}}]{Butler04}
{Butler}, R.~P., {Bedding}, T.~R., {Kjeldsen}, H., {et~al.} 2004, \apjl, 600,
  L75

\bibitem[{{Butler} {et~al.}(1996){Butler}, {Marcy}, {Williams}, {McCarthy},
  {Dosanjh}, \& {Vogt}}]{Butler96}
{Butler}, R.~P., {Marcy}, G.~W., {Williams}, E., {et~al.} 1996, \pasp, 108, 500

\bibitem[{{Carrier} {et~al.}(2007){Carrier}, {Kjeldsen}, {Bedding}, {Brewer},
  {Butler}, {Eggenberger}, {Grundahl}, {McCarthy}, {Retter}, \&
  {Tinney}}]{Carrier07}
{Carrier}, F., {Kjeldsen}, H., {Bedding}, T.~R., {et~al.} 2007, \aap, 470, 1059

\bibitem[{{Chaplin} {et~al.}(2011){Chaplin}, {Kjeldsen},
  {Christensen-Dalsgaard}, {Basu}, {Miglio}, {Appourchaux}, {Bedding},
  {Elsworth}, {Garc{\'{\i}}a}, {Gilliland}, {Girardi}, {Houdek}, {Karoff},
  {Kawaler}, {Metcalfe}, {Molenda-{\.Z}akowicz}, {Monteiro}, {Thompson},
  {Verner}, {Ballot}, {Bonanno}, {Brand{\~a}o}, {Broomhall}, {Bruntt},
  {Campante}, {Corsaro}, {Creevey}, {Do{\u g}an}, {Esch}, {Gai}, {Gaulme},
  {Hale}, {Handberg}, {Hekker}, {Huber}, {Jim{\'e}nez}, {Mathur}, {Mazumdar},
  {Mosser}, {New}, {Pinsonneault}, {Pricopi}, {Quirion}, {R{\'e}gulo},
  {Salabert}, {Serenelli}, {Aguirre}, {Sousa}, {Stello}, {Stevens}, {Suran},
  {Uytterhoeven}, {White}, {Borucki}, {Brown}, {Jenkins}, {Kinemuchi}, {Van
  Cleve}, \& {Klaus}}]{Chaplin11Sci}
{Chaplin}, W.~J., {Kjeldsen}, H., {Christensen-Dalsgaard}, J., {et~al.} 2011,
  Science, 332, 213

\bibitem[{{Christensen-Dalsgaard}(2004)}]{CD04}
{Christensen-Dalsgaard}, J. 2004, \solphys, 220, 137

\bibitem[{{Claudi} {et~al.}(2009){Claudi}, {Benatti}, {Bonanno}, {Bonavita},
  {Desidera}, {Gratton}, {Leccia}, {Cosentino}, {Endl}, \& {Carolo}}]{Claudi09}
{Claudi}, R., {Benatti}, S., {Bonanno}, A., {et~al.} 2009, Communications in
  Asteroseismology, 159, 21

\bibitem[{{Cox} \& {Pilachowski}(2000)}]{Cox2000}
{Cox}, A.~N. \& {Pilachowski}, C.~A. 2000, Physics Today, 53, 100000

\bibitem[{{Cox} \& {Smith}(1981)}]{Cox81}
{Cox}, J.~P. \& {Smith}, R.~C. 1981, The Observatory, 101, 87

\bibitem[{{Elsworth} \& {Thompson}(2004)}]{Elsworth04}
{Elsworth}, Y.~P. \& {Thompson}, M.~J. 2004, Astronomy and Geophysics, 45,
  050000

\bibitem[{{Flower}(1996)}]{Flower96}
{Flower}, P.~J. 1996, \apj, 469, 355

\bibitem[{{Frandsen} {et~al.}(1995){Frandsen}, {Jones}, {Kjeldsen}, {Viskum},
  {Hjorth}, {Andersen}, \& {Thomsen}}]{Frandsen95}
{Frandsen}, S., {Jones}, A., {Kjeldsen}, H., {et~al.} 1995, \aap, 301, 123

\bibitem[{{Fuhrmann}(2004)}]{Fuhrmann04}
{Fuhrmann}, K. 2004, Astronomische Nachrichten, 325, 3

\bibitem[{{Gilliland} {et~al.}(2010){Gilliland}, {Brown},
  {Christensen-Dalsgaard}, {Kjeldsen}, {Aerts}, {Appourchaux}, {Basu},
  {Bedding}, {Chaplin}, {Cunha}, {De Cat}, {De Ridder}, {Guzik}, {Handler},
  {Kawaler}, {Kiss}, {Kolenberg}, {Kurtz}, {Metcalfe}, {Monteiro}, {Szab{\'o}},
  {Arentoft}, {Balona}, {Debosscher}, {Elsworth}, {Quirion}, {Stello},
  {Su{\'a}rez}, {Borucki}, {Jenkins}, {Koch}, {Kondo}, {Latham}, {Rowe}, \&
  {Steffen}}]{Gilliland10a}
{Gilliland}, R.~L., {Brown}, T.~M., {Christensen-Dalsgaard}, J., {et~al.} 2010,
  \pasp, 122, 131

\bibitem[{{Gratton} {et~al.}(2001){Gratton}, {Bonanno}, {Bruno}, {Cali},
  {Claudi}, {Cosentino}, {Desidera}, {Diego}, {Farisato}, {Martorana},
  {Rebeschini}, \& {Scuderi}}]{Gratton01}
{Gratton}, R.~G., {Bonanno}, G., {Bruno}, P., {et~al.} 2001, Experimental
  Astronomy, 12, 107

\bibitem[{{Gray} {et~al.}(2006){Gray}, {Corbally}, {Garrison}, {McFadden},
  {Bubar}, {McGahee}, {O'Donoghue}, \& {Knox}}]{Gray06}
{Gray}, R.~O., {Corbally}, C.~J., {Garrison}, R.~F., {et~al.} 2006, \aj, 132,
  161

\bibitem[{{Grundahl} {et~al.}(2009){Grundahl}, {Christensen-Dalsgaard},
  {Kjeldsen}, {J{\o}rgensen}, {Arentoft}, {Frandsen}, \&
  {Kj{\ae}rgaard}}]{SONG}
{Grundahl}, F., {Christensen-Dalsgaard}, J., {Kjeldsen}, H., {et~al.} 2009, in
  Astronomical Society of the Pacific Conference Series, Vol. 416, Astronomical
  Society of the Pacific Conference Series, ed. {M.~Dikpati, T.~Arentoft,
  I.~Gonz{\'a}lez Hern{\'a}ndez, C.~Lindsey, \& F.~Hill}, 579

\bibitem[{{Kjeldsen} \& {Bedding}(1995)}]{Kjeldsen95}
{Kjeldsen}, H. \& {Bedding}, T.~R. 1995, \aap, 293, 87

\bibitem[{{Kjeldsen} {et~al.}(2008){Kjeldsen}, {Bedding}, {Arentoft}, {Butler},
  {Dall}, {Karoff}, {Kiss}, {Tinney}, \& {Chaplin}}]{Kjeldsen08baql}
{Kjeldsen}, H., {Bedding}, T.~R., {Arentoft}, T., {et~al.} 2008, \apj, 682,
  1370

\bibitem[{{Kjeldsen} {et~al.}(2005){Kjeldsen}, {Bedding}, {Butler},
  {Christensen-Dalsgaard}, {Kiss}, {McCarthy}, {Marcy}, {Tinney}, \&
  {Wright}}]{Kjeldsen05cenB}
{Kjeldsen}, H., {Bedding}, T.~R., {Butler}, R.~P., {et~al.} 2005, \apj, 635,
  1281

\bibitem[{{Kjeldsen} {et~al.}(1995){Kjeldsen}, {Bedding}, {Viskum}, \&
  {Frandsen}}]{Kjeldsen95boo}
{Kjeldsen}, H., {Bedding}, T.~R., {Viskum}, M., \& {Frandsen}, S. 1995, \aj,
  109, 1313

\bibitem[{{Kjeldsen} \& {Frandsen}(1992)}]{Kjeldsen92}
{Kjeldsen}, H. \& {Frandsen}, S. 1992, \pasp, 104, 413

\bibitem[{{Koch} {et~al.}(2010){Koch}, {Borucki}, {Basri}, {Batalha}, {Brown},
  {Caldwell}, {Christensen-Dalsgaard}, {Cochran}, {DeVore}, {Dunham},
  {Gautier}, {Geary}, {Gilliland}, {Gould}, {Jenkins}, {Kondo}, {Latham},
  {Lissauer}, {Marcy}, {Monet}, {Sasselov}, {Boss}, {Brownlee}, {Caldwell},
  {Dupree}, {Howell}, {Kjeldsen}, {Meibom}, {Morrison}, {Owen}, {Reitsema},
  {Tarter}, {Bryson}, {Dotson}, {Gazis}, {Haas}, {Kolodziejczak}, {Rowe}, {Van
  Cleve}, {Allen}, {Chandrasekaran}, {Clarke}, {Li}, {Quintana}, {Tenenbaum},
  {Twicken}, \& {Wu}}]{Koch10}
{Koch}, D.~G., {Borucki}, W.~J., {Basri}, G., {et~al.} 2010, \apjl, 713, L79

\bibitem[{{Leccia} {et~al.}(2007){Leccia}, {Kjeldsen}, {Bonanno}, {Claudi},
  {Ventura}, \& {Patern{\`o}}}]{Leccia07}
{Leccia}, S., {Kjeldsen}, H., {Bonanno}, A., {et~al.} 2007, \aap, 464, 1059

\bibitem[{{Luck} \& {Heiter}(2005)}]{Luck05}
{Luck}, R.~E. \& {Heiter}, U. 2005, \aj, 129, 1063

\bibitem[{{Marcy} \& {Butler}(1992)}]{Geoffrey92}
{Marcy}, G.~W. \& {Butler}, R.~P. 1992, \pasp, 104, 270

\bibitem[{{Mathur} {et~al.}(2011){Mathur}, {Handberg}, {Campante},
  {Garc{\'{\i}}a}, {Appourchaux}, {Bedding}, {Mosser}, {Chaplin}, {Ballot},
  {Benomar}, {Bonanno}, {Corsaro}, {Gaulme}, {Hekker}, {R{\'e}gulo},
  {Salabert}, {Verner}, {White}, {Brand{\~a}o}, {Creevey}, {Do{\u g}an},
  {Elsworth}, {Huber}, {Hale}, {Houdek}, {Karoff}, {Metcalfe},
  {Molenda-{\.Z}akowicz}, {Monteiro}, {Thompson}, {Christensen-Dalsgaard},
  {Gilliland}, {Kawaler}, {Kjeldsen}, {Quintana}, {Sanderfer}, \&
  {Seader}}]{Mathur11}
{Mathur}, S., {Handberg}, R., {Campante}, T.~L., {et~al.} 2011, \apj, 733, 95

\bibitem[{{Michel} {et~al.}(2008){Michel}, {Baglin}, {Auvergne}, {Catala},
  {Samadi}, {Baudin}, {Appourchaux}, {Barban}, {Weiss}, {Berthomieu},
  {Boumier}, {Dupret}, {Garcia}, {Fridlund}, {Garrido}, {Goupil}, {Kjeldsen},
  {Lebreton}, {Mosser}, {Grotsch-Noels}, {Janot-Pacheco}, {Provost},
  {Roxburgh}, {Thoul}, {Toutain}, {Tiph{\`e}ne}, {Turck-Chieze}, {Vauclair},
  {Vauclair}, {Aerts}, {Alecian}, {Ballot}, {Charpinet}, {Hubert},
  {Ligni{\`e}res}, {Mathias}, {Monteiro}, {Neiner}, {Poretti}, {Renan de
  Medeiros}, {Ribas}, {Rieutord}, {Cort{\'e}s}, \& {Zwintz}}]{CoRoT08}
{Michel}, E., {Baglin}, A., {Auvergne}, M., {et~al.} 2008, Science, 322, 558

\bibitem[{{Montgomery} \& {O'Donoghue}(1999)}]{Montgomery99}
{Montgomery}, M.~H. \& {O'Donoghue}, D. 1999, Delta Scuti Star Newsletter, 13,
  28

\bibitem[{{Nordgren} {et~al.}(1999){Nordgren}, {Germain}, {Benson},
  {Mozurkewich}, {Sudol}, {Elias}, {Hajian}, {White}, {Hutter}, {Johnston},
  {Gauss}, {Armstrong}, {Pauls}, \& {Rickard}}]{Nordgren1999}
{Nordgren}, T.~E., {Germain}, M.~E., {Benson}, J.~A., {et~al.} 1999, \aj, 118,
  3032

\bibitem[{{Oja}(1986)}]{Oja86}
{Oja}, T. 1986, \aaps, 65, 405

\bibitem[{{Quirion} {et~al.}(2010){Quirion}, {Christensen-Dalsgaard}, \&
  {Arentoft}}]{Quirion10}
{Quirion}, P., {Christensen-Dalsgaard}, J., \& {Arentoft}, T. 2010, \apj, 725,
  2176

\bibitem[{{Richichi} {et~al.}(2005){Richichi}, {Percheron}, \&
  {Khristoforova}}]{Richichi05}
{Richichi}, A., {Percheron}, I., \& {Khristoforova}, M. 2005, \aap, 431, 773

\bibitem[{{Stello} {et~al.}(2009){Stello}, {Chaplin}, {Basu}, {Elsworth}, \&
  {Bedding}}]{Stello09}
{Stello}, D., {Chaplin}, W.~J., {Basu}, S., {Elsworth}, Y., \& {Bedding}, T.~R.
  2009, \mnras, 400, L80

\bibitem[{{Tassoul}(1980)}]{Tassoul80}
{Tassoul}, M. 1980, \apjs, 43, 469

\bibitem[{{Valenti} \& {Fischer}(2005)}]{Valenti05}
{Valenti}, J.~A. \& {Fischer}, D.~A. 2005, \apjs, 159, 141

\bibitem[{{van Leeuwen}(2007)}]{HIPPARCOS}
{van Leeuwen}, F. 2007, \aap, 474, 653

\end{thebibliography}

\end{document}